# Full Nexus between Newtonian and Relativistic Mechanics

G. Sardin (gsardin@ub.edu)

*Universitat de Barcelona - Facultat de Física, Av. Diagonal, 628 - 08028 Barcelona*

A full nexus between Newtonian and relativistic mechanics is set. Contrarily to what is commonly thought, Newtonian mechanics can be amended to suit all speeds up to c. It is demonstrated that when introducing the fact that the pulse of oscillators, i.e. emitters and clocks, is sensitive to speed, the Newtonian framework can be extended to all speeds. For this aim, it is formulated the concept of actor scenario vs. observer scenario. This differentiation is essential to avoid confusion between effective reality (actor scenario) and appearance (observer scenario). Measurements are subjected to kinematical aberrations, the observer scenario being inertial. These must be removed to attain intrinsic reality, i.e. that of actors. The lack of demarcation between the two scenarios leads to conceptual confusions. The amended Newtonian mechanics is of full application. Here, it has been mainly applied to the Newtonian Doppler effect, amended to suit all speeds.

## I. Introduction

A full nexus between Newtonian [1] and relativistic [2] mechanics is established. Contrarily to what is commonly thought and taught in text books [3], Newtonian mechanics can be extended to high speeds up to that of light, by merely taking into account the fact that oscillators, i.e. clocks and emitters, are sensitive to speed. Newtonian mechanics is commonly presented as being an approximation that works out only for low speeds. This is so only because in the historical conventional presentation it is implicitly assumed that the pulse of clocks and emitters is independent of speed and thus taken as constant. In Newton times their dependence on speed was not suspected. However, nowadays we know they are and that their frequency slows down with increasing speed. So, let us include this fact into the Newtonian framework and remove the shortcoming of the standard Newtonian mechanics, making it so valid for any speed.

For this aim, let us express the fundamentals of the conceptual framework used, based on the concept of actor scenario and observer scenario. This differentiation is essential to avoid confusion about what corresponds to effective physical reality (actor scenario) and what is just appearance (observer scenario). Experimental results, which belong to the observer scenario, are subjected to kinematical aberrations due to the fact that the observer scenario is inertial. These must be removed to accede to intrinsic reality, i.e. that of actors. In classical mechanics, actors do behave independently of observers, and in fact they even do not perceive their eventual presence. So, actors are self-ruling and the physical laws to which they are subjected are certainly independent of the observer scenario. The absence of demarcation between the two scenarios leads to conceptual confusions, a burden of special relativity in which framework appearance and intrinsic reality cannot be differentiated. For example, the dilation of time, i.e. the slowing down of clock, is an actual physical effect while the contraction of space is a mathematically derived fictitious effect, but special relativity cannot make such distinction because it lacks of a proper conceptual framework that would avoid such coarse apprehension of physical reality.

As an application of the present approach we have formulated the Newtonian Doppler effect, amended to suit all speeds. The classical Doppler effect is conventionally presented to work out for sound waves and to fail for electromagnetic waves, for which it must be substituted by the relativistic formulation. This has become such a cliché that it is widely repeated without any chance of getting a deeper look. However, away from conventionalisms [4-7] it is easy to see that it is not imperious to jump into the relativistic standpoint [8-13]. It is enough to just improve the Newtonian premises by including the slowing down, fittingly to the $\gamma$ factor, of the frequency of oscillators as their speed increases. The invariance of space and the concept of absolute speed are preserved, as well as absolute time, i.e. that from absolute rest. These three notions belong to the actor scenario and express the intrinsic physical reality, since free of the kinematical aberrations derived from observations made on inertial systems, such as the Doppler effect related to the observer motion, which consequently belongs to the observer scenario. The amended Newtonian mechanics is of full application, and to illustrate it, it has also been briefly applied to the so-called twin paradox and to atmospheric muons.

## II. Developments

Let us first apply the amended Newtonian framework to the Doppler effect and consider a media in which a wave propagates at a speed $w$.

### II.1. Actor scenario

Let us address the case of an emitter (oscillator) moving at a speed $v_e$ with respect to an observer (detector). The frequency of the oscillator is at first considered to be independent of its motion, which applies for low speeds. However, the emitted wave is affected by the emitter motion since during a period it has shifted of a distance equal to $v_e\, T$. Consequently, the wavefronts are compacted or expanded in accordance to the respective direction of motion of the emitter and the wave. So, the wavelength is shortened or lengthened of an amount $\Delta\lambda = v_e\, T$ and hence $\lambda_d = \lambda \pm \Delta\lambda$, where $\lambda_d$ is the detected wavelength, $\lambda$ the actual wavelength and $T$ the period. This is not an apparent effect related to detection, instead it is a factual effect affecting the wavelength itself. Since this effect is related to the emitter it belongs to the actor scenario. The resultant detected frequency is thus:

$$f_d = \frac{w}{\lambda_d} = \frac{w}{\lambda \pm \Delta\lambda} = \frac{w}{(w \pm v_e)T} = f\frac{1}{(1 \pm \frac{v_e}{w})} \tag{1}$$

Let us envisage now the case of an emitter moving away, at absolute speed $v_e$, from an observer at rest with the media. For low speeds the emitted frequency is:

$$f(v_e) = f(0)\frac{1}{(1 + \frac{v_e}{w})} \tag{2}$$

where $f(v_e)$ is the wave frequency at the emitter absolute velocity $v_e$, $f(0)$ the corresponding frequency at absolute rest, and $w$ the propagation speed of the wave in the media.

If the emitter moves towards the observer, the emitted frequency is:

$$f(v_e) = f(0)\frac{1}{(1 - \frac{v_e}{w})} \tag{3}$$

In order to shorten the text, from now on only the case in which emitter and observer move apart will be considered. The inverse only implies a matter of sign.

These expressions are known not to suit for high speeds. The reason for it simply is that, in this classical formulation of the Doppler effect, it has been implicitly presumed that the emitter oscillation frequency is constant, an assumption that fails for high speeds. Actually, the departure of the classical formulation is due to the fact that the emitter is sensitive to speed, such that its oscillation frequency decreases as its speed through the media increases, leading to an increase of the wavelength of the emitted wave. So, let us reformulate the Doppler effect taking into account this fact.

*High-speed Doppler emendation related to the emitter:*

A high-speed correction has to be accounted for, which is related to the dependence of the proper frequency of the oscillator on its own speed, which in turn rebounds on the wavelength of the emitted wave. As before, this is not an apparent effect but it is instead a factual effect affecting the wavelength itself of the emitted wave. So, when accounting for the dependence of the emitted wave frequency on the emitter own speed, its formulation for high speeds becomes:

$$f(v_e) = f(0) \frac{\sqrt{1 - \left(\frac{v_e}{w}\right)^2}}{\left(1 + \frac{v_e}{w}\right)} = f(0) \sqrt{\frac{\left(1 - \frac{v_e}{w}\right)\left(1 + \frac{v_e}{w}\right)}{\left(1 + \frac{v_e}{w}\right)^2}} = f(0) \sqrt{\frac{\left(1 - \frac{v_e}{w}\right)}{\left(1 + \frac{v_e}{w}\right)}} \qquad (4)$$

where $v_e$ is the absolute speed of the emitter.

## II.2. Observer scenario

Depending on the observer motion being forwards or backwards the incoming wave at a speed $v_o$ its apparent speed will thus be $w \pm v_o$ and hence the frequency of impinging wavefronts will vary such that:

$$f_d = \frac{w \pm v_o}{\lambda} = \frac{f}{w}(w \pm v_o) = f(1 \pm \frac{v_o}{w}) \qquad (5)$$

where $f_d$ is the detected frequency, $\lambda$ the actual wavelength, $f$ the actual frequency and $w$ the wave velocity.

Let us now considerer the case of an observer (detector), moving away at an absolute speed $v_o$, from an emitter (oscillator) at rest with respect to the media. The detected apparent frequency is:

$$f(v_o) = f(0)(1 - \frac{v_o}{w}) \qquad (6)$$

where $f(v_o)$ is the frequency perceived by the observer with absolute speed $v_o$, $f(0)$ is the frequency of the emitted wave for the emitter being at absolute rest, $w$ is its propagation velocity within the media, and the term in parenthesis is the Doppler factor. Let us stress that the Doppler effect induced by the motion of the inertial observer is a kinematical effect exclusively related to perception, which does not affect the wave itself but only the way it is perceived.

Again this expression is known not to suit for high speeds. This time, the reason for it is that, in this classical formulation of the Doppler effect, it has been assumed that the pulse of the observer clock is constant, a presumption that also fails for high speeds. Actually, the observer clock is sensitive to speed, such that its frequency decreases as its speed through the media increases, leading to what is mathematically regarded as time dilation. So, let us reformulate the Doppler effect taking into account this fact.

*High-speed Doppler emendation related to the observer:*

The clock measuring the incoming wave frequency is itself affected by its own absolute speed (which is that of the inertial observer), in a way that its rhythm lowers with increasing speed. This is not a kinematical effect proceeding from the motion of the inertial observer. This frequency slowing down is a kinetic effect acting directly on the clock beating, and consequently affects the measure of the incoming wave frequency, but it does not affect the wave itself. It is thus an apparent effect affecting only the perception of the incoming wave frequency. So, when accounting for the dependence of the beating of the clock on its own speed, the expression of the wave frequency for high speeds becomes:

$$f(v_o) = f(0) \frac{1 - \frac{v_o}{w}}{\sqrt{1 - \left(\frac{v_o}{w}\right)^2}} = f(0) \sqrt{\frac{\left(1 - \frac{v_o}{w}\right)^2}{\left(1 - \frac{v_o}{w}\right)\left(1 + \frac{v_o}{w}\right)}} = f(0) \sqrt{\frac{\left(1 - \frac{v_o}{w}\right)}{\left(1 + \frac{v_o}{w}\right)}} \qquad (7)$$

### II.3. Observer scenario + actor scenario

Let us now considerer the case of an emitter and an observer moving away from each other, with respective absolute speed $v_e$ and $v_o$ relative to the media. In that case, the resultant apparent frequency according to the classical expression of the Doppler effect is:

$$f(v_e, v_o) = f(0) \frac{\left(1 - \frac{v_o}{w}\right)}{\left(1 + \frac{v_e}{w}\right)} \qquad (8)$$

*High-speed Doppler emendation related to the emitter and observer motions:*

When taking into account the observer scenario and the actor scenario, i.e. when the emitter and the observer are moving at high speed, the corrective factors from their motion have both to be accounted for, so the resultant expression of the detected frequency is:

$$f(v_e, v_o) = f(0) \frac{\left(1 - \frac{v_o}{w}\right)\sqrt{1 - \left(\frac{v_e}{w}\right)^2}}{\left(1 + \frac{v_e}{w}\right)\sqrt{1 - \left(\frac{v_o}{w}\right)^2}} = f(0) \sqrt{\frac{\left(1 - \frac{v_o}{w}\right)^2 \left(1 - \frac{v_e}{w}\right)\left(1 + \frac{v_e}{w}\right)}{\left(1 + \frac{v_e}{w}\right)^2 \left(1 - \frac{v_o}{w}\right)\left(1 + \frac{v_o}{w}\right)}} \qquad (9)$$

$$f(v_e, v_o) = f(0) \sqrt{\frac{\left(1 - \frac{v_e}{w}\right)\left(1 - \frac{v_o}{w}\right)}{\left(1 + \frac{v_e}{w}\right)\left(1 + \frac{v_o}{w}\right)}} \qquad (10)$$

which is the accurate Newtonian formulation of the detected frequency for any speed, expressed in absolute value. Let us point out that it has been obtained by distinguishing between the effect of speed on the measurement of the wave frequency, which corresponds to the classical Doppler effect, and the effect of speed on the frequency of the emitter and the clock themselves, a necessary amendment for high speeds.

**III. Mathematical nexus with special relativity**

The general formulism developed for the amended Newtonian framework, in which w stands for any speed, can be extended to electromagnetism by just taking the propagation velocity of the wave equal to c. Furthermore it allows a mathematical nexus with special relativity. In effect, the relativistic expression for the Doppler effect when emitter and observer move away from each other at a speed $u$, is:

$$f_d = f_e \sqrt{\frac{\left(1 - \frac{u}{c}\right)}{\left(1 + \frac{u}{c}\right)}} \tag{11}$$

where $f_d$ is the detected frequency and $f_e$ is the actual emitted frequency.

Hence, it suffices to establish equality between equations (10) and (11), i.e. to define a match between the Doppler effect expressed in the absolute speeds $v_e$ and $v_o$ of the emitter and observer and in their relative speed u. The terms within the square roots (Doppler factors) should thus be equal:

$$\frac{\left(1 - \frac{v_e}{w}\right)\left(1 - \frac{v_o}{w}\right)}{\left(1 + \frac{v_e}{w}\right)\left(1 + \frac{v_o}{w}\right)} = \frac{\left(1 - \frac{u}{c}\right)}{\left(1 + \frac{u}{c}\right)} \tag{12}$$

In the case the emitter and the observer would move towards each other, equation (11) would be:

$$\frac{\left(1 + \frac{v_e}{w}\right)\left(1 + \frac{v_o}{w}\right)}{\left(1 - \frac{v_e}{w}\right)\left(1 - \frac{v_o}{w}\right)} = \frac{\left(1 + \frac{u}{c}\right)}{\left(1 - \frac{u}{c}\right)} \tag{13}$$

For both equations (12) and (13), equality is fulfilled for the same value of u:

$$u = \frac{v_e + v_o}{1 + \frac{v_e v_o}{c^2}} \tag{14}$$

These relations constitute a nexus between absolute and relative speeds, and hence as well, a bridge to special relativity, at the same time the above formulations based on amended grounds of Newtonian mechanics are ascertained.

**IV. Application to the twin paradox**

Within the strict framework of special relativity the twin paradox has no solution since the principle of relativity does not allow it. In effect, for each twin it is the other one who moves away from him at high speed and therefore, once getting back together, each one expects to be the other one who has aged less. The impossibility to break down this symmetry without appealing to any artificial justification or to any widget outside the frame of special relativity, forbids getting a solution to this presumed paradox. In contrast, within the amended Newtonian framework there is no such shortcoming since there is no such symmetry, all speeds being expressed with respect to a single reference. So, each twin ages according to his own absolute speed, and their speed difference defines their aging variance. Here, there is no indistinctness about who would age slower since being unambiguously the one who has travelled at the higher absolute speed, or that is to say, at higher speed in regard to the Cosmic Microwave Background.

$$t_1 = \frac{t_0}{\sqrt{1-\left(\frac{v_1}{c}\right)^2}} \quad \text{and} \quad t_2 = \frac{t_0}{\sqrt{1-\left(\frac{v_2}{c}\right)^2}} \tag{15}$$

$$\Delta t = t_2 - t_1 = \frac{t_0}{\sqrt{1-\left(\frac{v_2}{c}\right)^2}} - \frac{t_0}{\sqrt{1-\left(\frac{v_1}{c}\right)^2}} \tag{16}$$

where $\Delta t$ is the aging difference between the twins, $t_1$ and $t_2$ refer to the time indicated by each twin clock, $t_0$ is the elapsed time at absolute rest, and $v_1$ and $v_2$ are the twins absolute speed.

**V. Application to atmospheric muons**

The speed-induced dilation of time is:

$$\tau = \frac{\tau_0}{\sqrt{1-\left(\frac{v}{c}\right)^2}} \tag{17}$$

where $\tau_0$ is the intrinsic life-time of muons and $v$ their absolute speed.

The amended Newtonian mechanics through the slowing down of oscillators (generically called dilation of time) does not need any space contraction to explain how atmospheric muons reach the Earth ground. In order to account for this fact special relativity must appeal to the contraction of space since from the point of view of muons it is the Earth that is, at once, speeding towards them at diverse high speeds (one for each muon), in view that each one takes itself as reference.

$$d = d_0 \sqrt{1-\left(\frac{v}{c}\right)^2} \tag{18}$$

where $d$ is the distance from the muons point of view and $d_0$ the actual distance measured on Earth, and $v$ stands for the relative speeds between muons and the Earth.

Since Newtonian mechanics does not use the proper system as a referential frame, and refers instead all speeds to a single referential frame embodied by a media or space itself, it only needs the so-called dilation of time. Therefore, in the Newtonian framework the unique cause allowing muons to reach the Earth surface is the dilation of their lifetime along with their absolute speed. The contraction of space is not needed, which stresses that it is just a fictitious effect, only required by a given math as a consequence of its conceptual fundamentals. It is just a mathematical wildcard, specific to special relativity and having no physical counterpart.

$$\tau_m = \frac{\tau_0}{\sqrt{1-\left(\frac{v_m}{c}\right)^2}} \quad \text{and} \quad \tau_T = \frac{\tau_0}{\sqrt{1-\left(\frac{v_T}{c}\right)^2}} \tag{19}$$

$$\Delta t = \tau_m - \tau_T = \tau_T = \frac{\tau_0}{\sqrt{1-\left(\frac{v_m}{c}\right)^2}} - \frac{\tau_0}{\sqrt{1-\left(\frac{v_T}{c}\right)^2}} \tag{20}$$

where $\tau_m$ is the life-time of muons at an absolute speed $v_m$, $t_T$ is the time dilation on Earth due to its net absolute speed $v_T$, which is very low (of only 370 km/s) [14], and $v_m$ that of the atmospheric muons, which is close to that of light. So, on practical grounds the Earth velocity can be neglected and muons reach the Earth ground due to the dilation of their mean lifetime derived from their high absolute speed.

## VI. Conclusion

It has been shown that when one takes into account the fact that any oscillator is sensitive to its absolute speed, the Newtonian mechanics passes to give correct results for all speeds. Indeed, this is so because oscillators, whatever emitters or clocks, are governed by absolute speeds and do not address relative velocities. Moreover, there is no need to appeal to any alleged contraction of space. From this, it arises a reflection of deep philosophical meaning, i.e. it has been obtained an accurate Newtonian formulation that does not require the manipulation of space. This indicates that its assumed contraction does not correspond to any actual physical reality, and that it is just a need of a specific math. This highlights that mathematical extrapolations must not be transposed in an animist-like way to physical reality. The mislead transcription of the Lorentz transformation to physical reality has led to regard space contraction as an actual physical effect instead of just as a mathematical contrivance. By only taking into account the sensitivity of oscillators to their speed through space it is enough to give proper consideration to the effects induced by high speeds.

We have expressed the Doppler effect in absolute value for any speed up to c, and established a link between the Newtonian mechanics, based on absolute speeds, and special relativity, which is restricted to relative velocities. We believe that the apprehension of kinetic phenomena upwind to the sole relative velocities has impoverished physics, at the time it has led to a conception of physical reality that relies on a counterfeit causality appealing to fictitious effects, such as the alleged contraction of space. Let us also stress that the theory of relativity requires the contraction of space to be a factual physical effect, without which it cannot e.g. account for the fact that muons created in the atmosphere reach the Earth surface despite their too short intrinsic lifetime does not allow it. The contraction of space is a mathematical requisite restricted to the theory of relativity that derives from counterfeit fundamentals.

In Newton's times the dependence of the oscillation frequency on speed was not suspected, and so, neither the variation of the emission frequency nor the measure of time. Likewise, the relativistic approach could be accepted by the times Einstein put it, given the failure of the detection of the presumed ether. What cannot be understood is its maintenance, due to relativist doctrinal attachments, after the discovery of the cosmic radiation background, which has far overcome the hypothetical ether [13], since it can be seen as an electromagnetic ether, offering so the advantage of being an energetic media easily detectable. Indeed, the cosmic background [14] constitutes an absolute referential frame, despite its dogmatic rebuff from relativists who put the doctrine above experimental evidences, since composed of photons flying in all directions at speed c, and hence its net motion is necessarily null. So, due to its photonic nature the cosmic microwave background cannot have any drift velocity within space, and being highly isotropic it constitutes a media which provides an ideal frame of reference, furthermore of universal extension. Absolute rest is determined when its blackbody spectrum provides a minimal absolute temperature.

The slowing down of oscillators frequency (clocks and emitters) with increasing speed is an intrinsic property which does not derive from observation on an inertial system, contrarily to the Doppler effect due to the observer motion, which is an extrinsic effect not affecting the wave itself but only its perception. It is essential to differentiate between intrinsic reality and appearance, i.e. between actor scenario and observer scenario. The mathematical workability of a theory does not necessarily make it a good theory on conceptual grounds, nor the best approach. The fact is that special relativity, by denying absolute speed and so being limited to relative velocities, constitutes a needlessly amputated approach to reality. Absolute speeds coexist with relative speeds, and to deny the formers just impoverishes physics, since absolute speeds allow to get relative speeds but not the inverse.

The framework used applies the factor $\gamma$ to the emitter and to the clock, using their absolute speed, while special relativity only applies it to the clock, since merely using their relative speed. Doing so, information is lost in the relativity framework, and this deteriorates the access to causality. In negating absolute speeds, and so relying exclusively on relative speeds, the theory of special relativity has hindered the access to physical reality. The present approach based on the amended Newtonian mechanics can be generalised to the whole kinetics. It ends up that physical reality is described through the amended Newtonian mechanics much more straightforwardly and comprehensively than through special relativity.

**References**

[1] Newton I., *The Principia: Mathematical Principles of Natural Philosophy*, Univ. of California Press Publ. (1999)

[2] Einstein, A., *Relativity*, Crown Publishers, N. Y. (1988)

[3] Tipler, P.A., *Physics*, Worth Pub., New York (1981)

[4] Sardin G., Animated Presentation: *Cause of the mechanical and electromagnetic equivalency of inertial systems*, http://inertial.equivalency.googlepages.com/animatedpresentation (2008)

[5] Sardin G., *A causal approach to first-order optical equivalency of inertial systems, by means of a beampointing test-experiment based on speed-induced deflection of light*, Physics e-Print archive: http://arxiv.org/pdf/physics/0401091 (2004)

[6] Sardin G., *First and second order electromagnetic equivalency of inertial systems, based on the wavelength and the period as speed-dependant units of length and time*, Physics e-Print archive: http://arxiv.org/pdf/physics/0401092 (2004)

[7] Sardin G., *Testing Lorentz symmetry of special relativity by means of the Virgo or Ligo set-up, through the differential measure of the two orthogonal beams time-of-flight*, Physics e-Print archive: http://arxiv.org/pdf/physics/0404116 (2004)

[8] Terletskii, Y., *Paradoxes in the Theory of Relativity*, Plenium Press Ed., New York (1968)

[9] Brush S.G., *Why was Relativity accepted?*, Phys. Perspect., 1 (1999) pp.184-214

[10] Roberton H.P., *Postulate versus Observation in the Special Theory of Relativity*, Rev. Mod. Phys., 21 (1949) pp.378-382

[11] Bohm D., *The Special Theory of Relativity*, Ed. W.A. Benjamin, Inc., New York (1965)

[12] Schleif S., *Experimental Basis of Special Relativity*, http://www2.corepower.com:8080/~relfaq/experiments.html

[13] Whittaker E.T., *A History of the Theories of Aether and Electricity*, AIP Publ. (1987)

[14] Cosmic Microwave Background: Speed relative to CMB, Editor: Wikipedia, http://en.wikipedia.org/wiki/CMB